\documentclass[10pt,conference]{IEEEtran}

\IEEEoverridecommandlockouts
\usepackage{cite}
\usepackage{amsmath,amssymb,amsfonts}
\usepackage{algorithmic}
\usepackage{graphicx}
\usepackage{textcomp}
\usepackage[table]{xcolor} 
\usepackage{physics}
\usepackage{booktabs}
\usepackage{multirow}
\usepackage[flushleft]{threeparttable}
\usepackage{colortbl}
\usepackage{siunitx}

\AtBeginDocument{\RenewCommandCopy\qty\SI}
\ExplSyntaxOn
\msg_redirect_name:nnn { siunitx } { physics-pkg } { none }
\ExplSyntaxOff

\usepackage{algorithm}
\usepackage{rotating}
\usepackage{capt-of}
\usepackage{hyperref}
\usepackage[activate={true,nocompatibility},final,tracking=true,kerning=true,spacing=true,factor=1100,stretch=10,shrink=10]{microtype}

\def\BibTeX{{\rm B\kern-.05em{\sc i\kern-.025em b}\kern-.08em
    T\kern-.1667em\lower.7ex\hbox{E}\kern-.125emX}}
\begin{document}
\bstctlcite{IEEEexample:BSTcontrol}

\title{Scalable Transpilation for Overcoming Restricted Connectivity in Distributed Superconducting Quantum Architectures

\thanks{This work was supported by the IQST Graduate School @QuantumBW, funded by the Ministry of Science, Research and Arts of Baden-Württemberg.}
}

\author{
\IEEEauthorblockN{Afonso Azenha}
\IEEEauthorblockA{\textit{Institut für Technische Informatik and} \\
\textit{Center for Integrated Quantum} \\ 
\textit{Science and Technology (IQST)} \\
\textit{University of Stuttgart} \\
Stuttgart, Germany \\
afonso.azenha@iti.uni-stuttgart.de}
\and
\IEEEauthorblockN{Ilia Polian}
\IEEEauthorblockA{\textit{Institut für Technische Informatik and} \\
\textit{Center for Integrated Quantum} \\ 
\textit{Science and Technology (IQST)} \\
\textit{University of Stuttgart} \\
Stuttgart, Germany \\
ilia.polian@iti.uni-stuttgart.de}
\and
\IEEEauthorblockN{Sebastian Brandhofer}
\IEEEauthorblockA{\textit{IBM Quantum, IBM Research Europe} \\
Ehningen, Germany \\
sebastian.brandhofer2@ibm.com}
}

\IEEEaftertitletext{\vspace{-2.25em}}
\maketitle

\begin{abstract}
Distributed Quantum Computing (DQC) offers a promising pathway to scale superconducting quantum hardware, but restricted inter- and intra-module connectivity creates a major transpilation bottleneck. Current DQC transpilation methods often assume idealized hardware connectivity, ignore subsequent intra-module routing, or suffer from substantial computational overheads that limit their applicability at scale. To address this, we introduce a scalable transpilation method tailored for the restricted connectivity of distributed superconducting quantum architectures. Specifically, our quantum interaction graph (QIG) approach couples a graph-partitioning mapping step with a fast SABRE-inspired routing heuristic. On a $48$-qubit ($3~\text{QPUs} \times 16~\text{qubits}$) architecture with all-to-all inter-core connectivity, our method reduces average aggregated costs by up to $24.72\%$ (unstructured) and $24.01\%$ (structured), while delivering $>250\times$ speed-ups for structured workloads compared to the state-of-the-art DMapS algorithm. On the $399$-qubit ($3\times133$) IBM Flamingo, where DMapS times out on dense circuits, our approach maintains practical runtimes. Against a DQC-adapted SABRE baseline, it reduces structured circuit routing costs by up to $51.94\%$ (all-to-all inter-core topology) and $51.96\%$ (restricted line topology). Ultimately, this work delivers a practical transpilation method for DQC at scale.
\end{abstract}

\begin{IEEEkeywords}
    Distributed Quantum Computing, Quantum Circuit Transpilation, Hypergraph Partitioning, Graph Partitioning, Hardware-Aware Routing, Superconducting Architectures
\end{IEEEkeywords}

\vspace{-0.25em}
\section{Introduction}
\label{SecI:Introduction}
Quantum computing promises transformative capabilities for problems that challenge the limits of classical computation across diverse application domains, including quantum chemistry \cite{Cao_2019, McArdle_2020}, materials science \cite{Bauer_2020}, and combinatorial optimization \cite{Moll_2018}. However, engineering the large-scale quantum processors necessary to unlock these capabilities presents immense challenges on both the hardware and software fronts~\cite{bravyi2022future}.

While superconducting platforms have achieved impressive scaling milestones \cite{ibm2025nighthawk,google2024willow, kim2023evidence, google2023suppressing}, the traditional monolithic approach of housing all qubits on a single chip faces escalating difficulties. As qubit counts grow, severe bottlenecks emerge, including highly complex control electronics, demanding cryogenic cooling power requirements, and sharply decreasing manufacturing yields \cite{heya2025randomized}.

To overcome these limitations, Distributed Quantum Computing (DQC) offers a compelling, scalable alternative \cite{caleffi2024distributed}. This paradigm interconnects multiple smaller quantum processors (referred to as quantum processing units (QPUs) or \textit{cores} interchangeably in this work) via remote quantum links to form a unified system, typically via quantum teleportation mediated by Einstein-Podolsky-Rosen (EPR) pairs \cite{Kozlowski_2019}. This architectural shift bypasses the aforementioned constraints of monolithic chips and is only now becoming physically viable through recent advances in quantum networking that enable these remote links (see, e.g., \cite{schär2026modularcryogeniclinkmicrowave}). However, this remains an active area of research, with current link fidelities about an order of magnitude lower than those of local (intra-QPU) gates \cite{heya2025randomized, caleffi2024distributed}. This stark discrepancy makes the efficient use of EPR pairs an absolute necessity for practical DQC in the pre-fault-tolerant era and beyond.

Despite the availability of advanced transpilation algorithms for monolithic processors \cite{li2019tacklingqubitmappingproblem, murali2019noiseadaptive, niu2020hardwareaware, ferrari2022noiseadaptive, liu2022swapscostcaseoptimizationaware, zou2024lightsabrelightweightenhancedsabre}, applying them directly to DQC environments often yields suboptimal performance. These algorithms naturally lack the context to penalize costly remote communications or navigate restricted inter-core topologies. Conversely, current DQC-specific literature tends to narrowly focus on the initial mapping problem (assigning virtual qubits to physical cores), frequently treating the subsequent routing phase (inserting SWAP gates) as a secondary concern \cite{andresmartinez2019automated, baker2020fgproee, wu2022autocomm, mao2023mhsa, bandic2023qubo, wu2023entanglementefficient,escofet2023hungarian, andresmartinez2024pytketdqc, escofet2025hqa}. In reality, transpiling for DQC involves navigating a complex optimization landscape governed by the interplay between inter-QPU EPR pair usage and local SWAP gate insertions. Notably, optimizing solely to minimize EPR pair usage can inadvertently trigger a massive surge in local SWAPs. In such scenarios, the accumulated errors from excessive local routing may ultimately outweigh the benefits of conserving EPR pairs, despite the higher error rates of the remote links themselves. This highlights the critical need for end-to-end transpilation methods capable of actively balancing these potentially competing overheads.

Compounding these challenges, existing research commonly relies on idealized hardware assumptions, such as all-to-all connectivity both within and between cores, that fail to reflect realistic distributed superconducting architectures. Literature addressing joint mapping and routing under strict topological constraints remains remarkably sparse. For instance, while open-source tools like pytket-dqc \cite{andresmartinez2024pytketdqc} exist, they focus primarily on mapping and lack native, integrated intra-core routing. Conversely, although holistic methodologies have been proposed \cite{ferrari2023modular, luo2025dmaps}, a lack of open-source implementations leaves DMapS \cite{luo2025dmaps} as the state-of-the-art end-to-end baseline used in this work. However, while DMapS improves upon prior techniques, its computational overhead grows significantly faster than that of lightweight heuristics (such as SABRE~\cite{li2019tacklingqubitmappingproblem}-inspired algorithms) as circuits become larger and denser. Ultimately, the steep computational overhead of existing state-of-the-art methodologies (including DMapS and pytket-dqc) renders them increasingly impractical for near-term utility-scale architectures, requiring the development of highly scalable transpilation workflows.

To address these limitations, we introduce hardware-aware transpilation heuristics specifically developed for connectivity-constrained DQC environments at scale. Rather than relying on a single algorithmic approach, we develop and analyze several customizable methodologies, ranging from lightweight heuristics and graph partitioning to more computationally intensive hypergraph-based techniques. Because transpilation efficiency and quality are intrinsically tied to the underlying circuit structure, providing multiple heuristic options maximizes the likelihood of identifying a highly performant strategy across a wide variety of quantum workloads.

The remainder of this paper is structured as follows. Sections \ref{SecII:Related_Work} and \ref{SecIII:Background} review the state-of-the-art and provide essential background on quantum circuit transpilation, the SABRE algorithm, and the DQC paradigm. Section \ref{SecIV:SABRE_Adaptations_for_DQC} details our SABRE adaptations for DQC, while Section \ref{SecV:Partitioning-Based_Mapping} introduces our partitioning-based approaches. Section \ref{SecVI:Evaluation} evaluates our algorithms against the state-of-the-art, and Section \ref{SecVII:Conclusions_and_Future_Work} discusses conclusions and future directions.

\section{Related Work}
\label{SecII:Related_Work}
Prior attempts to address the qubit mapping and routing problem in distributed quantum computing are broadly defined by their hardware assumptions and algorithmic approaches. Static partitioning methods, for instance, assign virtual qubits to physical cores in a single, unchanging initial layout. The Multi-stage Hybrid Simulated Annealing (MHSA) algorithm \cite{mao2023mhsa} optimizes this mapping to minimize remote communication by using a simulated annealing meta-heuristic algorithm. However, MHSA decouples global mapping from local routing, ignoring the associated local SWAP overhead. Addressing mapping through hypergraph partitioning \cite{andresmartinez2019automated} allows two-qubit gates to be grouped into packets for efficient multi-gate teleportation \cite{eisert2000ejpp}. This foundational idea was refined into pytket-dqc \cite{andresmartinez2024pytketdqc}, an extension for the tket compiler \cite{Sivarajah_2020} that maximizes EPR reuse through advanced techniques such as \textit{embeddings} \cite{wu2023entanglementefficient}, ``detached'' gates, and Steiner trees. Yet, this produces large hypergraphs whose subsequent post-processing introduces substantial computational overhead as circuits grow.

Alternatively, dynamic mapping methods continuously re-optimize qubit allocations across discrete time slices (in the quantum circuit). Approaches such as FGP-rOEE \cite{baker2020fgproee} achieve this by partitioning modified interaction graphs (where edges represent two-qubit gates and weights confer a degree of lookahead) to dynamically assign virtual qubits to cores for each time-slice. The Hungarian qubit assignment (HQA) \cite{escofet2023hungarian, escofet2025hqa} subsequently advanced this paradigm by utilizing the Hungarian algorithm \cite{Kuhn1955} to dynamically assign gates, rather than qubits, to specific QPUs. While time-sliced assignments suit fully connected architectures (e.g., trapped ions), they do not directly translate to superconducting platforms featuring sparse coupling maps. By outputting only qubit-to-core assignments rather than exact physical placements, these methods abstract away the required intra-core SWAP insertions and precisely scheduled teleportations, leaving the true routing overheads unaccounted for.

Recognizing the necessity of integrated mapping and routing, recent studies have shifted toward end-to-end transpilation. Ferrari et al. \cite{ferrari2023modular} propose a modular framework that minimizes resource consumption and circuit depth by integrating qubit assignment, local routing, and a scheduling strategy utilizing quantum state and gate teleportation primitives, while considering both network and device constraints. As a publicly available, fully realized pipeline, DMapS \cite{luo2025dmaps} achieves end-to-end transpilation for superconducting hardware via a two-stage approach. It establishes an initial layout using a simplified hypergraph model, followed by the heuristic assignment of virtual to physical qubits, before employing complex routing heuristics to prioritize intra-QPU over inter-QPU SWAP insertions. However, these intricate SWAP-selection heuristics introduce a steep increase in computational overhead when routing dense circuits at scale. This overhead directly motivates the lightweight, SABRE-inspired heuristics and quantum interaction graph (QIG) strategies proposed in this work.

\section{Background}
\label{SecIII:Background}
To execute abstract quantum circuits on physical hardware, transpilers must resolve limited connectivity constraints. The physical layout of these connections is formally represented by a coupling map, an undirected graph where nodes correspond to physical qubits and edges denote allowable two-qubit interactions. Resolving these constraints involves mapping (assigning virtual to physical qubits) and routing (inserting SWAP gates to satisfy the coupling map's connectivity).

\subsection{SABRE Algorithm}
\label{SubSecIII:SABRE}
The SwAp-based BidiREctional heuristic search algorithm (SABRE) \cite{li2019tacklingqubitmappingproblem} is a leading mapping and routing method originally designed for monolithic QPUs. It serves as the standard in Qiskit \cite{javadiabhari2024quantumcomputingqiskit} via an optimized implementation known as LightSABRE \cite{zou2024lightsabrelightweightenhancedsabre}.

SABRE represents a quantum circuit as a directed acyclic graph (DAG) where nodes represent gates, classical bits, and virtual qubits, and edges denote dependencies. The algorithm tracks a front layer ($F$) of gates ready for immediate execution and an extended set ($E$) of direct descendants acting as a lookahead window. Starting from a random initial mapping, SABRE iteratively routes the circuit by evaluating candidate SWAP insertions. A candidate SWAP exchanges the mapped virtual states of two adjacent physical qubits. 

To select the best SWAP, SABRE computes a heuristic cost evaluating the physical distance between the target qubits ($q_1$ and $q_2$) for each unresolved two-qubit gate $g$. Gates in $F$ are weighted more heavily to address pressing routing needs, while $E$ provides capacity to anticipate future operations. Once a SWAP makes target qubits adjacent, the corresponding gate executes, and $F$ and $E$ are updated with the next unresolved gates from the DAG. To escape poor initializations, SABRE performs bidirectional routing, alternating forward and backward passes. Specifically, the final physical layout of a forward routing pass serves as the initial layout for a subsequent backward pass on the reversed circuit, helping converge on a high-quality initial layout before the final forward routing pass.

SABRE's routing decisions rely on a distance matrix derived from the hardware coupling map. The algorithm natively assigns a distance of 1 to adjacent physical qubits and computes shortest paths between all pairs using the Floyd-Warshall algorithm. Crucially, modifying this matrix to incorporate hardware characteristics (e.g., error rates) directly tunes SABRE's performance. Furthermore, the heuristic cost function itself is customizable.

To evaluate a candidate SWAP, SABRE computes the cost of the resulting virtual-to-physical mapping $\pi$ using the lookahead heuristic cost in (\ref{eq:sabre_lookahead_heuristic}), where $D$ represents the distance matrix and $W$ dictates the influence of the lookahead set $E$ relative to the front layer $F$ (Qiskit typically applies $W = 0.5$).

\vspace{-0.5em}
\begin{equation}
    \label{eq:sabre_lookahead_heuristic}
    \begin{split}
        H_{lookahead} & = \sum_{g \in F} D[\pi(g.q_1)][\pi(g.q_2)] \\
        & + \frac{W}{|E|} \sum_{g \in E} D[\pi(g.q_1)][\pi(g.q_2)]
    \end{split}
\end{equation}
We follow Qiskit in leaving the front-layer term unnormalized; the original SABRE~\cite{li2019tacklingqubitmappingproblem} averages it over $|F|$.

\subsection{Transpilation for Distributed Quantum Computing}
\label{SubSecIII:DQC}
The DQC paradigm scales computational power by interconnecting multiple QPUs. We model this network as an undirected graph $G = (C, L)$, where the vertex set $C$ represents the individual cores and the edge set $L$ denotes the connecting quantum communication links. These links enable the transmission of quantum information between spatially separated cores via shared EPR pairs, $\ket{\Phi^+} = \frac{1}{\sqrt{2}} (\ket{00} + \ket{11})$. By consuming shared EPR pairs to perform quantum gate or state teleportation, distant QPUs can execute inter-core operations. This allows the network to function as a unified quantum processor, albeit one where remote links incur substantially higher costs than local connections.

Gate teleportation executes a remote two-qubit gate across two QPUs without permanently transferring either computational state between the cores. For example, a remote Controlled-Z (CZ) requires one EPR pair alongside local operations and classical communication. Consequently, executing a remote SWAP via its standard three-CZ decomposition requires three EPR pairs. Alternatively, state teleportation moves a virtual qubit's state entirely to a target QPU; this can achieve a remote SWAP at the cost of two EPR pairs, but requires the quantum link to support two simultaneous pairs between the communicating cores. Furthermore, multi-gate teleportation protocols like the Eisert-Jacobs-Papadopoulos-Plenio (EJPP) protocol \cite{eisert2000ejpp} allow a single EPR pair to mediate a sequence of remote gates. This capability forms the foundation for hypergraph-based mapping methods that group gates into distributable packets \cite{andresmartinez2024pytketdqc}.

To improve transpilation efficiency in DQC environments, an ideal integrated approach requires a mechanism to incorporate local routing costs directly into the global mapping stage. By representing the distributed architecture as a single conjoined coupling map that includes both intra-core physical connections and inter-core quantum links, modified versions of the SABRE heuristic could leverage bidirectional routing passes to co-optimize local SWAP insertions and remote EPR consumption simultaneously.

\section{SABRE Adaptations for DQC}
\label{SecIV:SABRE_Adaptations_for_DQC}

To adapt quantum circuit transpilation for distributed architectures and avoid the sub-optimal routing of monolithic approaches, we introduce several modifications to the SABRE heuristic. By customizing the underlying coupling map, distance matrix, heuristic cost function, and the selection criteria for the best trial, we extend the framework to natively account for DQC-specific constraints. This approach retains SABRE's lightweight scalability while dynamically penalizing the high cost of non-local gates, establishing a more DQC-aware routing foundation.

\subsection{Conjoined Coupling Map}
\label{SubSecIV:Conjoined_Map}

\begin{figure*}[t]
    \centering
    \includegraphics[width=0.80\textwidth]{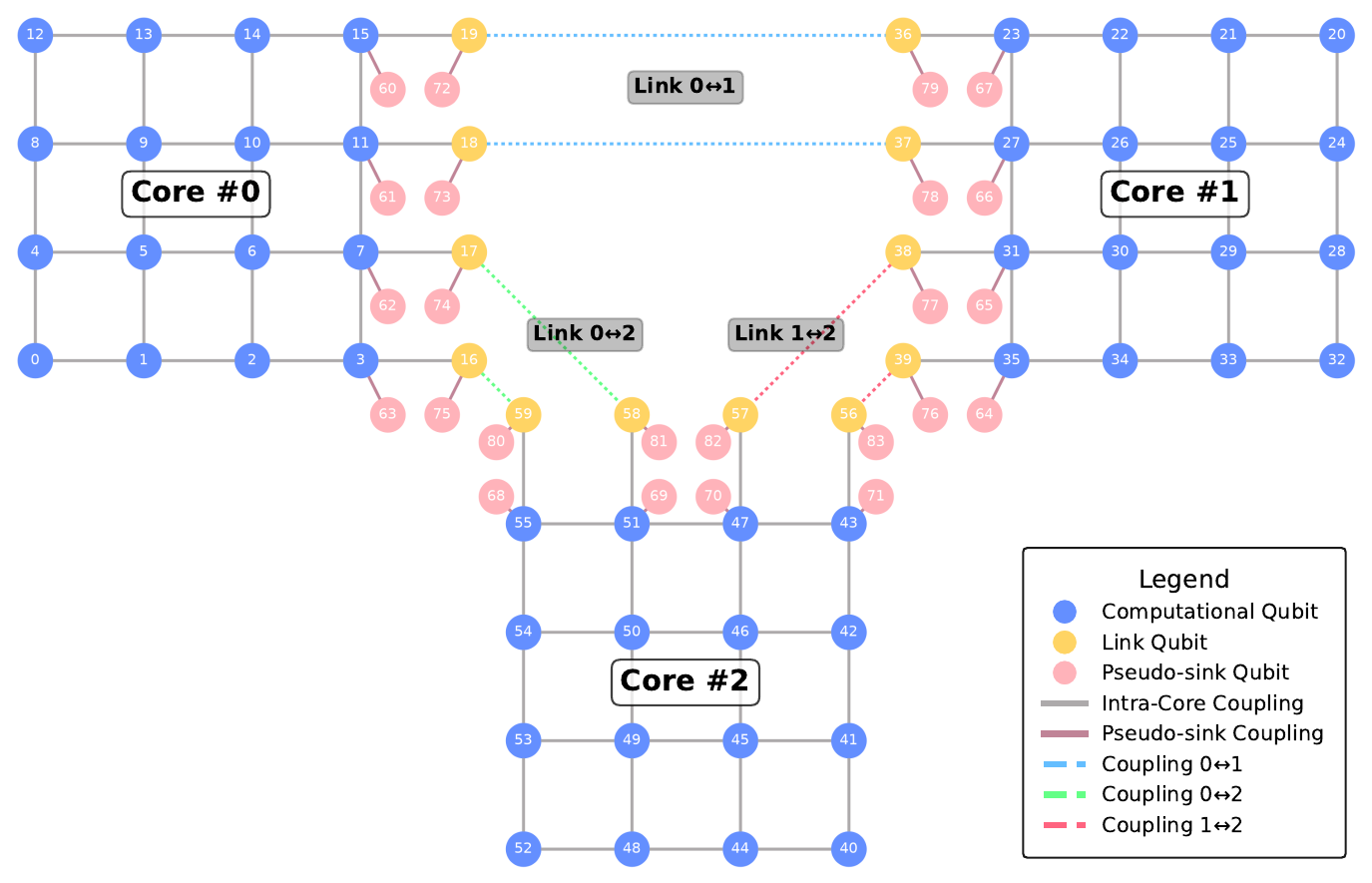}
    \caption{Illustration of a conjoined coupling map for a $3\times16$ distributed architecture. Each core's local coupling map is represented as a cluster of computational qubits (blue nodes) with square connectivity. In addition to these computational qubits, dedicated link qubits (yellow nodes) are explicitly modeled in our hypergraph-based methods to handle inter-core communication. These link qubits serve as the physical terminals for the distribution of EPR pairs and mediate the sequential quantum gate teleportation protocols. Intra-core physical connections are depicted as solid gray edges, while remote quantum links are represented by the colored dashed lines connecting the cores. Auxiliary pseudo-sink qubits (pink nodes) and their associated solid edges are also specific to these hypergraph methods; while not physically part of the hardware, these abstract nodes serve as routing anchors to force computational qubits into the communication zone during pytket-dqc's subcircuit routing (detailed in Section \ref{SubSecV:pytket_dqc}).}
    \label{fig:conjoined_map}
    \vspace{-1.5em}
\end{figure*}

Monolithic transpilers route circuits over standalone coupling maps. For DQC, we instead construct a \textit{conjoined coupling map}. This structure stitches the individual cores' coupling graphs together by introducing additional edges representing the remote quantum links (Fig. \ref{fig:conjoined_map}).

We assume a realistic near-term scenario where each remote link sustains a maximum capacity of a single shared EPR pair (i.e., $1$ ebit) at any given time \cite{heya2025randomized}.\footnote{For hypergraph-based methods, this constraint implies allocating a single pair of dedicated physical link qubits per remote connection (see Fig. \ref{fig:conjoined_map}).} This restriction limits remote operations to sequential gate teleportation, precluding state teleportation protocols that would require two simultaneous ebits of memory. Because a standard SABRE implementation is blind to these underlying entanglement costs, it natively treats a costly inter-core SWAP exactly the same as a local SWAP, often resulting in sub-optimal routing outcomes that over-utilize remote links.

\subsection{Distance Matrix Customization: (1, 10) SABRE}
\label{SubSecIV:1_10_SABRE}

To prevent the over-insertion of remote SWAPs, we manipulate the edge weights of the underlying conjoined coupling map before computing the all-pairs distance matrix. We introduce a SABRE routing variant denoted as \textbf{$(1, 10)$ SABRE}. In this model, intra-QPU edges are assigned a weight of 10, while inter-QPU edges are assigned a weight of 1. While assigning a higher weight to the ``cheaper'' local edges seems counter-intuitive, it fundamentally exploits SABRE's greedy logic. In SABRE, a SWAP is chosen based on how much it reduces the summed distances between the interacting qubits of the pending gates in $F$ and $E$. By increasing the local edge weights, a local SWAP is perceived by the algorithm as a high-reward move. Consequently, the algorithm is more likely to exhaust local routing pathways before resorting to remote links, though an inter-QPU SWAP can still be selected if the gates within the extended set $E$ strongly incentivize it.

It is worth noting that there are parallels between this scheme and error-aware transpilation methods (e.g., \cite{murali2019noiseadaptive, niu2020hardwareaware, ferrari2022noiseadaptive}), where distance matrices, and occasionally cost functions, are modified to reflect individual link fidelities. However, much like our $(1, 10)$ SABRE heuristic, these methods do not natively account for the complexities of distributed architectures. Because they lack a mechanism for explicit EPR pair counting or minimization, we anticipate that achieving highly optimized inter-core communication likely requires more sophisticated formulations that integrate DQC-specific information. To address this limitation, we explore more elaborate methodologies in the subsequent sections.

Beyond routing logic, we modify SABRE's initial layout selection criteria. While standard parallel SABRE trials select the run with the fewest SWAP gates inserted, we instead prioritize the lowest \textit{aggregated cost} $C_{agg}$:
\begin{equation}
    \label{eq:aggregated_cost}
    C_{agg} = (W_{EPR} \times N_{EPR}) + N_{\text{Local\_SWAP}}, \vspace{-0.25em}
\end{equation}
where $N_{EPR}$ is the total EPR pair consumption and $N_{\text{Local\_SWAP}}$ is the number of local SWAPs inserted. We set the tunable penalty weight to $W_{EPR} = 10$, reflecting the current order-of-magnitude fidelity difference between remote and local operations \cite{heya2025randomized}. As distributed hardware evolves, this parameter can be dynamically adjusted to match the exact error rates of the target architecture.

\vspace{-0.25em}
\subsection{Cost Function Customization: CLA-SABRE}
\label{SubSecIV:CLA_SABRE}

While manipulating the distance matrix passively biases SABRE against remote SWAP insertions, we also propose actively modifying its heuristic cost function to explicitly recognize and reward multi-gate-teleportation-like circuit structures. 

We introduce \textbf{Custom Lookahead SABRE (CLA-SABRE)}, which modifies the standard lookahead heuristic (\ref{eq:sabre_lookahead_heuristic}) to dynamically penalize inter-core SWAPs unless they ``localize'' future remote gates (i.e., ensure the gates' interacting qubits reside in the same core) or reduce inter-core distances across enough pending gates. Conceptually, this adds an absolute penalty $CLA$ to the evaluation of candidate SWAP $S_j$:
\begin{equation}
\begin{split}
    H_{CLA} & = \sum_{g \in F} D[\pi(g.q_1)][\pi(g.q_2)] \\
    & + \frac{W}{|E|} \sum_{g \in E} D[\pi(g.q_1)][\pi(g.q_2)] + CLA.
\end{split}
\end{equation}

Following LightSABRE's relative scoring mechanism~\cite{zou2024lightsabrelightweightenhancedsabre}, we define only the SWAP-specific differential penalty, $\Delta H_{CLA}(S_j)$. For a local SWAP candidate $S_j = (u, v)$, $\Delta H_{CLA}(S_j) = 0$. If $S_j$ spans a remote link, it incurs a constant upfront penalty $\alpha$, which can be offset based on its impact on future pending gates:
\begin{equation}
\label{eq:cla_term}
    \Delta H_{CLA}(S_j) = \alpha + \sum_{p \in \{u, v\}} \sum_{g_p \in E_p} \phi(p, \bar{p}, k_{g_p}).
\end{equation}
Here, $E_p \subseteq E$ is the subset of pending gates involving the virtual qubit currently on physical qubit $p$, $\bar{p}$ represents its destination after the SWAP (meaning the pair $(p, \bar{p})$ corresponds to either $(u, v)$ or $(v, u)$ from $S_j$), and $k_{g_p}$ is the physical qubit hosting the \textit{other} operand of gate $g_p$.

To quantify the reward $\phi$, let $D_C(x, y)$ represent the inter-core distance between the host cores of physical qubits $x$ and $y$. This metric is derived from the DQC network topology, evaluating shortest paths strictly at the core level while abstracting away specific intra-core positions. We evaluate the current inter-core distance $d_{\text{curr}} = D_C(p, k_{g_p})$ and the post-SWAP distance $d_{\text{new}} = D_C(\bar{p}, k_{g_p})$. The resulting contribution $\phi$ is calculated as:
\begin{equation}
    \phi(p, \bar{p}, k_{g_p}) = 
    \begin{cases} 
    -\frac{\alpha}{3}, & \text{if } 0 = d_{\text{new}} < d_{\text{curr}}, \\
    \frac{\alpha}{3\beta} (d_{\text{new}} - d_{\text{curr}}), & \text{otherwise}.
    \end{cases}
\end{equation}

This formulation directly reflects EPR overheads, where an inter-core SWAP consumes three EPR pairs. Unlike the lookahead term in (\ref{eq:sabre_lookahead_heuristic}), neither the upfront penalty $\alpha$ nor the custom rewards ($\phi$) are scaled by $W/|E|$, so their magnitude does not shrink as the lookahead window grows. Thus, a per-gate reward of $-\frac{\alpha}{3}$ cancels the penalty $\alpha$ if a move localizes exactly three remote gates ($0 = d_{\text{new}} < d_{\text{curr}}$); localizing more yields a net reward. Conversely, moves that merely reduce inter-core distance earn partial rewards scaled by the distance reduction and dampened by $\beta$. We empirically set $\alpha = 9.0$ and $\beta = 3.0$. The evaluation runtime scales with $|E|$; we use the Qiskit standard $|E| = 20$ but also explore $|E| = 100$ to assess a deeper lookahead window.

\section{Partitioning-Based Mapping}
\label{SecV:Partitioning-Based_Mapping}

While our SABRE adaptations optimize routing on a conjoined map, an alternative paradigm utilizes graph and hypergraph partitioning. By solving the initial qubit-to-core assignment prior to routing, these methods inject explicit DQC structural awareness into the transpilation pipeline, reducing EPR consumption. In this section, we detail our adaptations of a state-of-the-art hypergraph mapping tool, followed by our partitioning-based methodologies.

\subsection{Hypergraph Partitioning}
\label{SubSecV:pytket_dqc}

The state-of-the-art hypergraph-based mapping tool, pytket-dqc \cite{andresmartinez2024pytketdqc}, minimizes EPR consumption by modeling the circuit as a hypergraph where nodes represent virtual qubits and two-qubit gates, and hyperedges encapsulate distributable packets. To maximize packet size, it employs \textit{embeddings} (extending packets across intermediate gates), Steiner trees (for EPR reuse), and ``detached'' gates (allowing operations to execute outside their assigned host cores).

Despite its sophistication, pytket-dqc assumes all-to-all intra-core connectivity, a model invalid for near-term superconducting hardware. Adapting it required major modifications to its tracking of remote physical link utilization and subcircuit generation logic. In EJPP-style multi-gate teleportation protocols, the starting and ending processes of a remote sequence are temporally separated, allowing local operations to occur during the intervening period. Under pytket-dqc's assumed connectivity, once the sender's link qubit is measured during the starting process, the framework assumes the now-free link qubit can be immediately reused to establish a new EPR pair with any other free link qubit. Consequently, pytket-dqc natively tracks only link qubit usage.

However, architectures with limited connectivity physically constrain this procedure. Because superconducting cores possess sparse intra-core connectivity, dedicated hardware nodes must be utilized as physical link qubits bound to specific remote links (see Fig. \ref{fig:conjoined_map}). During a multi-gate teleportation protocol, the transmitted quantum state arrives at the receiver's physical link qubit to participate in subsequent local gates. In the meantime, intra-core routing might temporarily move this active computational state out of the physical link qubit and into the computational grid, requiring it to be routed back later to finalize the teleportation primitive. Prematurely generating a new EPR pair on this specific remote link before the primitive concludes would irreversibly overwrite the physical link qubit, destroying any active computational state it holds. Consequently, we rewrote the distributed circuit generation logic to strictly track physical link usage, locking the link until the multi-gate teleportation primitive is fully resolved.

The second modification introduces explicit subcircuit generation for each core. Remote communications require virtual qubits to be physically routed to the immediate vicinity of the physical link qubits (the communication zone). To enforce this, we introduce auxiliary \textit{pseudo-sink qubits} (Fig. \ref{fig:conjoined_map}). For each scheduled remote operation, we insert a \textit{routing placeholder} (a dummy two-qubit gate) between the designated virtual qubit and the pseudo-sink qubit corresponding to the desired link. When these subcircuits are later routed in parallel, these placeholders force the virtual qubits into the communication zone at the exact time the teleportation is required.

To resolve these placeholders without accidentally moving active computational states onto restricted link or pseudo-sink qubits, we assign hierarchical edge weights to the coupling map (Table \ref{tab:edge_weights}) before computing the distance matrix. These large weight disparities, combined with an explicit heuristic penalty against SWAPs involving pseudo-sinks, freeze the sinks in place and safely route the required states. Once routing concludes, the auxiliary pseudo-sinks and routing placeholders are cleanly discarded. Conceptually, this mirrors the Big-M method in operations research \cite{winston2004operations}, where arbitrarily large costs are injected into an evaluation metric to strictly enforce logical constraints without altering the algorithm's underlying search mechanics.

\begin{table}[ht]
    \centering
    \caption{Custom Edge Weights for Intra-Core SABRE Routing}
    \vspace{-0.15cm}
    \label{tab:edge_weights}
    \begin{tabular}{@{}lcc@{}}
        \toprule
        \textbf{Edge Type} & \textbf{Weight} & \textbf{Routing Priority} \\ \midrule
        Computational $\leftrightarrow$ Computational & $10^9$ & $1^{\text{st}}$ (Highest) \\
        Link $\leftrightarrow$ Computational & $10^6$ & $2^{\text{nd}}$ \\
        Pseudo-sink $\leftrightarrow$ Computational & $10^3$ & $3^{\text{rd}}$ \\
        Pseudo-sink $\leftrightarrow$ Link & $10^0$ & $4^{\text{th}}$ (Lowest) \\ \bottomrule
    \end{tabular}
    \vspace{-1.75em}
\end{table}

Finally, we evaluate two distinct pytket-dqc mapping variants, both of which pair the initial hypergraph partitioning with specialized refinements: \texttt{PartitioningHeterogeneous} (PH), which adjusts partition boundaries to smooth communication costs, and \texttt{CoverEmbedding} (CE), a runtime-intensive extension formulating embedding optimization as a Vertex Cover problem. While effective, the severe computational overhead of these hypergraph analyses limits their scalability for large, dense circuits, motivating our subsequent methodologies.

\subsection{Hybrid: Hypergraph Partitioning and CLA-SABRE} 
\label{SubSecV:Hybrid_Mapping}

To bridge the gap between sophisticated hypergraph-based approaches and scalable routing heuristics, we introduce a hybrid methodology. To bypass the substantial computational overhead of constructing the distributed circuit, we utilize pytket-dqc's \texttt{PartitioningHeterogeneous} routine strictly to determine the initial qubit-to-core assignment. Once established, we immediately halt the pytket-dqc pipeline, skipping its expensive link-tracking and teleportation scheduling routines. Instead, we generate a random initial layout on the conjoined coupling map that strictly respects this core assignment.

We then feed the unrouted circuit and this constrained layout into CLA-SABRE. By explicitly rewarding multi-gate-teleportation-like structures during physical routing, CLA-SABRE utilizes its standard forward-backward passes to refine the global conjoined initial mapping. Ultimately, decoupling the initial core assignment from the physical routing phase allows this hybrid approach to extract the structural benefits of hypergraph analysis while bypassing pytket-dqc's runtime-intensive distributed circuit generation procedures. While this decoupling significantly reduces transpilation time, the internal refinement passes of \texttt{PartitioningHeterogeneous} still incur prohibitive runtimes for large, dense circuits, demanding a fundamentally different approach.

\subsection{Scalable Graph Partitioning (QIG + SABRE)}
\label{SubSecV:Scalable_QIG_Partitioning}

To bypass the hybrid approach's computational bottlenecks, we adapt established graph partitioning techniques utilized in recent DQC mapping schemes \cite{baker2020fgproee, bandic2023qubo}. However, rather than extracting time-sliced graphs, we construct a single, global quantum interaction graph (QIG). In the QIG, vertices represent virtual qubits, and edge weights denote the total frequency of two-qubit gates between any pair across the entire circuit. This abstraction makes the partitioning complexity dependent solely on the virtual qubit count rather than gate count, unlike hypergraph-based formulations.

We partition this lightweight QIG into distinct blocks (one per physical core) using the KaHyPar library \cite{schlag2022kahypar}, also employed by pytket-dqc, to ensure a common partitioning backend across methods. However, KaHyPar's default objective function minimizes the connectivity metric (equivalent to cut size), implicitly assuming an all-to-all inter-core topology. Because near-term DQC architectures typically lack this full connectivity, such a naive minimization is insufficient. To address this, we evaluate all block-to-core permutations, selecting the assignment that minimizes actual communication costs based on the target hardware topology.

Furthermore, while KaHyPar supports heterogeneous core capacities, the non-zero imbalance parameter ($\epsilon$) required for efficient solving cannot guarantee strict physical hardware limits per block. We address this, and further reduce communication costs, via a two-stage post-processing refinement:
\textbf{1) Strict Capacity Enforcement:} If a core is over-allocated, we evaluate its assigned virtual qubits, iteratively ejecting the candidate that incurs the minimum communication penalty (i.e., the smallest increase in remote two-qubit gates) upon relocation, and moving it to an under-allocated core until the hardware limit is respected. 
\textbf{2) Boundary Reallocation:} Next, we apply a cost-reduction pass inspired by pytket-dqc's namesake greedy refinement routine \cite{andresmartinez2024pytketdqc}. We iteratively evaluate virtual qubits on partition boundaries, shifting them to neighboring under-allocated cores if the move strictly reduces global communication costs. Because this strategy moves individual qubits into empty slots rather than swapping pairs across boundaries, its effectiveness naturally diminishes as the circuit size approaches the architecture's maximum capacity.

This computationally efficient initial mapping is then injected directly into our SABRE variants for physical routing (retaining standard forward-backward passes). Specifically, we pair this QIG mapping with distinct SABRE routing variants: Default SABRE, $(1, 10)$ SABRE, and CLA-SABRE (testing both the standard $|E|=20$ and an expanded $|E|=100$ lookahead window). Crucially, all evaluated SABRE variants are DQC-adapted implementations; as detailed in Section \ref{SecIV:SABRE_Adaptations_for_DQC}, they leverage our conjoined coupling map and modified layout selection criteria. Consequently, in our subsequent results, the ``Default SABRE'' baseline simply denotes the variant that utilizes these two underlying DQC adaptations but otherwise retains the standard SABRE routing logic.

Unlike the hypergraph formulation, a standard graph lacks explicit awareness of multi-gate teleportation structures. However, this QIG formulation excels at rapidly grouping highly interacting virtual qubits onto the same core. By deferring the localization of remote gates to the routing phase (e.g., via CLA-SABRE), this approach drastically reduces computational overhead while still yielding EPR consumption rates comparable to hypergraph methods.

\vspace{-0.25em}
\section{Evaluation}
\label{SecVI:Evaluation}
\vspace{-0.25em}

In this section, we evaluate the developed algorithms against the state-of-the-art across three distributed quantum architectures and various benchmark circuits. The code is available at \url{https://github.com/kaiuki2000/scalable_dqc_transpilation}.

\vspace{-0.40em}
\subsection{Experimental Setup}
\vspace{-0.10em}

\subsubsection{Hardware and Software Environment}
All evaluations were performed on an AMD Ryzen 7 PRO 6850U processor with 26 GB of RAM, running Ubuntu 24.04. To ensure reproducibility, the DMapS baseline utilized its official implementation (running on Qiskit 0.39.2). Our methodologies leverage Qiskit 2.2.0, with custom SABRE modifications implemented directly in Rust within Qiskit's compiled backend to ensure optimal execution speeds. Graph and hypergraph partitioning relies on KaHyPar 1.3.2, while hypergraph adaptations (Section \ref{SecV:Partitioning-Based_Mapping}) were built via source modifications to pytket-dqc. We enforced timeouts of 1 hour (48-qubit architecture) and 3 hours (399-qubit architectures). Runtimes reflect isolated, single-threaded execution averaged over five random seeds. Unless otherwise stated, all SABRE-based variants utilize $|E| = 20$.

\subsubsection{Distributed Quantum Architectures}
\label{SubSubSecVI:Distributed_Architectures}
We benchmark across three distributed architectures to evaluate scalability and topological awareness:

\begin{itemize}
    \item \textbf{48-Qubit ($3 \times 16$ square cores, all-to-all inter-core connectivity):} A controlled case study enabling direct comparison with DMapS and pytket-dqc before scalability bottlenecks become prohibitive. The physical layout corresponds to the architecture depicted in Fig. \ref{fig:conjoined_map}.
    \item \textbf{399-Qubit IBM Flamingo \cite{ibm2022qdc, ibm2024qdc} ($3 \times 133$ IBM Heron r1 cores, all-to-all inter-core connectivity):} A large-scale system modeling three 133-qubit quantum processors with full inter-core connectivity. This topology stress-tests our SABRE-inspired and QIG methods on dense circuits where DMapS and pytket-dqc encounter severe scalability bottlenecks.
    \item \textbf{399-Qubit IBM Flamingo ($3 \times 133$ IBM Heron r1 cores, line inter-core connectivity):} A constrained variant of the previous architecture, created by removing two remote links to enforce a strict inter-core line topology. This evaluates generalizability to heterogeneous networks.
\end{itemize}

\subsubsection{Benchmarking Circuits and Preprocessing}
We compile a diverse dataset of quantum circuits from standard suites (QASMBench \cite{qasmbench}, MQTBench \cite{mqtbench}, and Benchpress \cite{benchpress}), divided into structured and unstructured categories.

The \textbf{structured circuits} category features distinct algorithmic patterns, including standard textbook algorithms (e.g., QFT, Shor's, GHZ), near-term variational applications (e.g., VQE, QAOA), and Hamiltonian simulation circuits \cite{hamlib2024} (e.g., Transverse-Field Ising, Fermi-Hubbard, and Heisenberg models). Because certain routing heuristics may align more effectively with specific structural patterns, evaluating this diverse suite should allow us to observe preliminary correlations between circuit structure and heuristic performance.

Conversely, \textbf{unstructured circuits} lack predictable interaction patterns, pushing inter-core routing capabilities to their limits. We primarily utilize randomized ``CZ-fraction'' circuits, inspired by the evaluation methodology in \cite{andresmartinez2024pytketdqc}: per layer, a Hadamard gate is applied to each qubit with probability $1-p$, while the remaining unselected qubits are randomly paired for CZ operations. We generate 36-qubit circuits (36 layers) for the 48-qubit architecture and 399-qubit circuits (399 layers) for the IBM Flamingo topologies. To complement these, we include a Quantum Volume (QV) circuit for each architecture to evaluate performance on dense, unstructured workloads.

To capture the complex routing dynamics of large-scale DQC, we strictly select circuits that saturate at least two full cores (e.g., $\geq32$ qubits for the 48-qubit architecture), resulting in 20 structured benchmarks for the small topology and 16 for the larger ones. All selected circuits then undergo identical preprocessing: Qiskit's \texttt{transpile} (optimization level 2) resolves custom gates and reduces initial gate counts, followed by pytket-dqc's \texttt{DQCPass}, which translates the circuits into the restricted \{H, RZ, CU1\} basis natively required by the hypergraph approaches.

\subsection{Results and Discussion}

To formalize our evaluation, we define two primary average metrics. \textbf{Cost Reduction} is calculated as $(1 - \text{GM}(C_{\text{method}} / C_{\text{baseline}})) \times 100\%$, where $C$ is the aggregated routing cost (\ref{eq:aggregated_cost}) and $\text{GM}$ denotes the geometric mean across the respective structured or unstructured benchmark suite. A negative cost reduction indicates worse performance relative to the reference baseline. \textbf{Speed-up} is correspondingly calculated as the geometric mean of the runtime ratios ($T_{\text{baseline}} / T_{\text{method}}$). While space constraints preclude reporting EPR pair and local SWAP counts individually, we note that on the 399-qubit architectures EPR trends closely track the aggregated cost trends reported below, whereas on the 48-qubit architecture the aggregated cost is at times dominated by local SWAP insertions rather than EPR usage.

Our central objective is to demonstrate that the proposed QIG and SABRE-inspired methods achieve highly scalable transpilation for utility-scale hardware. Specifically, we aim to show they deliver competitive, and often superior, mapping and routing quality compared to state-of-the-art tools, while operating orders of magnitude faster. Average results are presented in Table \ref{tab:master_benchmark_results_a} (48-qubit) and Table \ref{tab:master_benchmark_results_b} (399-qubit), with circuit-level performance detailed in Figs. \ref{fig:48-qubits} and \ref{fig:399-qubits}. To maintain visual clarity, certain circuits are omitted from the bar plots, though all instances are fully accounted for in the averages reported in the tables.

\subsubsection{Performance and Scalability on the 48-Qubit Architecture}

% ============================================================
% TABLE II — 48-Qubit, single column (fits ~1 column width)
% ============================================================
\begin{table}[t]
    \centering
    \caption{
        Aggregated Cost Reduction and Runtime Speed-up (Geometric-Mean Ratios): 48-Qubit (3\,$\times$\,16, All-to-All), Relative to DMapS. Bold values are best per column. See footnotes for partial-result superscripts.         \textbf{Legend:}
        \textnormal{\itshape Def.\ SABRE} = DQC-adapted SABRE baseline;
        \textnormal{\itshape CLA} = Custom Lookahead SABRE;
        \textnormal{\itshape CE} = \texttt{CoverEmbedding};
        \textnormal{\itshape PH} = \texttt{PartitioningHeterogeneous};
        \textnormal{\itshape HGP} = (Hybrid) Hypergraph Partitioning;
        \textnormal{\itshape QIG} = Quantum Interaction Graph;
        \textbf{T.O.} = 1\,h (48-q) timeout.
    }
    \label{tab:master_benchmark_results_a}
    \vspace{-0.15cm}
    \resizebox{\linewidth}{!}{%
        \begin{tabular}{l rr p{0.5em} rr}
            \toprule
            \multirow{2}{*}{\textbf{Algorithm}}
                & \multicolumn{2}{c}{Unstructured ($7{\times}5{+}1$)}
                &
                & \multicolumn{2}{c}{Structured (20 Circuits)} \\
            \cmidrule(lr){2-3}\cmidrule(lr){5-6}
                & \multicolumn{1}{c}{Cost Red.}
                & \multicolumn{1}{c}{Speed-up}
                &
                & \multicolumn{1}{c}{Cost Red.}
                & \multicolumn{1}{c}{Speed-up} \\
            \midrule

            % --- SABRE family ---
            Def.\ SABRE
                & $-14.05\%$   & $\mathbf{428.449\times}$
                &
                & $-17.87\%$   & $328.991\times$ \\
            (1,10) SABRE
                & $-0.54\%$    & $393.313\times$
                &
                & $-68.18\%$   & $296.675\times$ \\
            CLA ($|E|{=}20$)
                & $8.93\%$     & $245.031\times$
                &
                & $-21.61\%$   & $264.429\times$ \\
            CLA ($|E|{=}100$)
                & $7.54\%$     & $125.559\times$
                &
                & $-36.53\%$   & $74.640\times$ \\
            \midrule

            % --- Hypergraph family ---
            pytket (CE)+Def.\ SABRE\textsuperscript{a}
                & $\mathbf{33.05\%}$\textsuperscript{a} & $0.428\times$\textsuperscript{a}
                &
                & T.O.\textsuperscript{b}  & T.O.\textsuperscript{b} \\
            pytket (PH)+Def.\ SABRE\textsuperscript{a}
                & $23.53\%$\textsuperscript{a}  & $0.256\times$\textsuperscript{a}
                &
                & T.O.\textsuperscript{b}  & T.O.\textsuperscript{b} \\
            HGP+CLA ($|E|{=}20$)
                & $26.70\%$    & $24.348\times$
                &
                & $\mathbf{28.94\%}$\textsuperscript{c} & $19.152\times$\textsuperscript{c} \\
            \midrule

            % --- QIG family ---
            QIG+Def.\ SABRE
                & $-8.90\%$    & $281.934\times$
                &
                & $16.87\%$    & $\mathbf{403.934\times}$ \\
            QIG+(1,10) SABRE
                & $9.75\%$     & $192.773\times$
                &
                & $5.34\%$     & $291.622\times$ \\
            QIG+CLA ($|E|{=}20$)
                & $24.72\%$    & $139.597\times$
                &
                & $24.01\%$    & $252.574\times$ \\
            QIG+CLA ($|E|{=}100$)
                & $15.95\%$    & $70.055\times$
                &
                & $5.44\%$     & $71.020\times$ \\
            \bottomrule
        \end{tabular}
    }%

    \vspace{0.15cm}
    \raggedright
    \scriptsize
    \textsuperscript{a} Subset (35/36); 1-hour timeout on Quantum Volume (32 qubits).\\
    \textsuperscript{b} \textbf{T.O.}: 1-hour timeout on 14/20 circuits; suite averages omitted due to unrepresentative sample size.\\
    \textsuperscript{c} Subset (19/20); 1-hour timeout on Shor's algorithm (42 qubits).
    \vspace{-1.75em}
\end{table}

% ============================================================
% FIG. 2 — 48-Qubit, portrait figure* (full width)
% ============================================================
\begin{figure*}[t]
    \centering
    \includegraphics[width=0.95\textwidth]{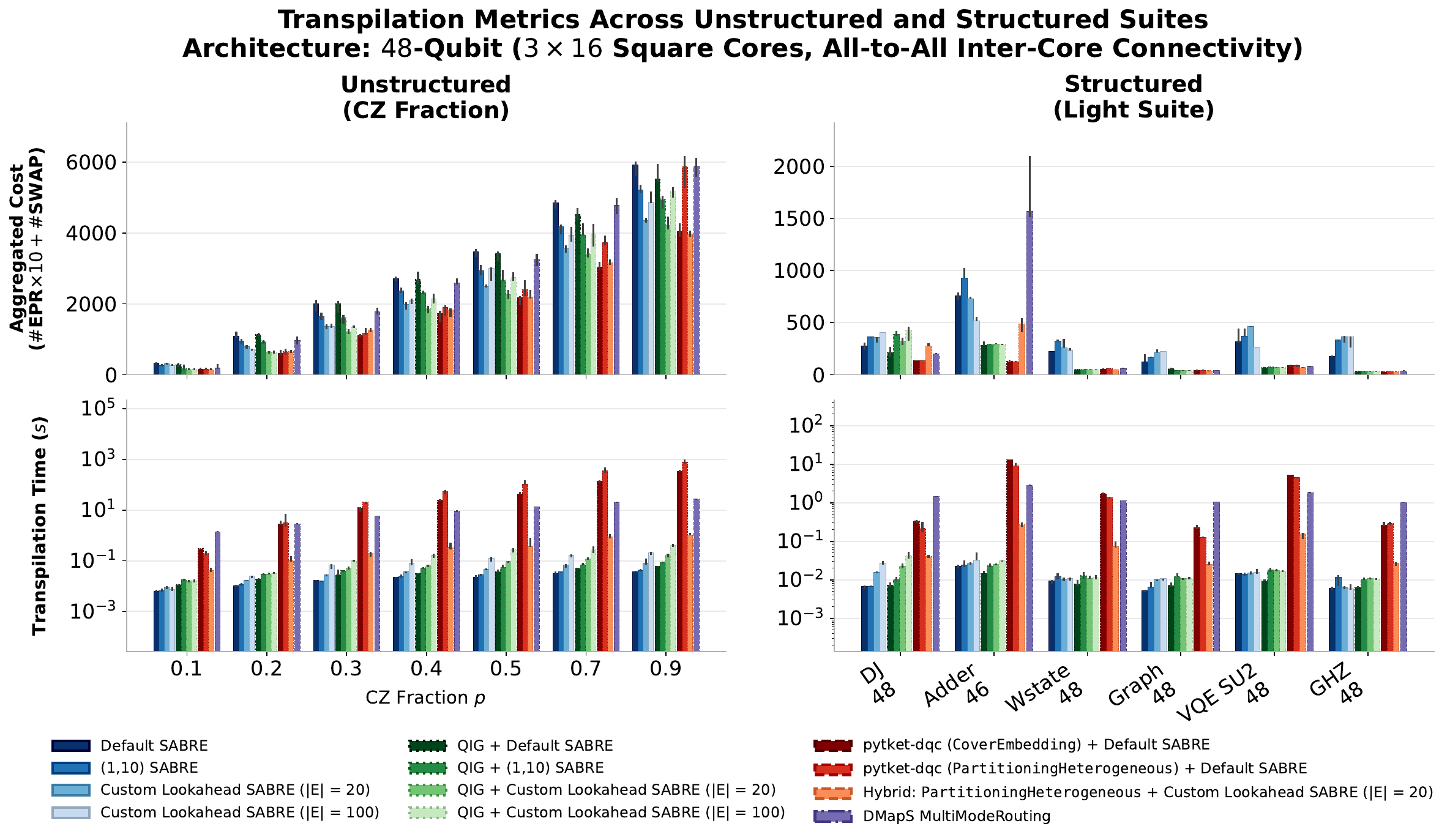}
    \vspace{-0.55em}
    \caption{Results for the 48-qubit architecture (all-to-all). Left column: unstructured circuits (CZ-fraction circuits); right column: ``light'' structured suite. Evaluated metrics: aggregated cost (\ref{eq:aggregated_cost}) and total transpilation time (1-hour timeout). Bars represent medians over 5 seeds; error bars show the central 75\% distribution (12.5th--87.5th percentiles). Color code: blues/greens for our proposed methods (SABRE/QIG variants); red/orange tones for hypergraph approaches (pytket-dqc, Hybrid); purple for DMapS. \textbf{Note:} ``Default SABRE'' is our DQC-adapted SABRE baseline (Sec.~\ref{SecIV:SABRE_Adaptations_for_DQC}).}
    \label{fig:48-qubits}
    \vspace{-1.50em}
\end{figure*}

We initially utilize DMapS as our reference baseline for the 48-qubit topology, as it represents the current state-of-the-art for end-to-end superconducting DQC transpilation.

Focusing first on the \textbf{unstructured} suite, hypergraph-based methods yield the highest transpilation quality (Table \ref{tab:master_benchmark_results_a}). By explicitly grouping gates into multi-gate teleportation packets, the best-performing pytket-dqc variant achieves a 33.05\% cost reduction over DMapS, though it operates at a mere $0.428\times$ the speed of DMapS. As demonstrated by its subsequent timeouts on the structured suite, this mathematically rigorous formulation leads to severe scaling limitations. Conversely, our Hybrid approach (HGP + CLA-SABRE) actively targets this scalability issue; it yields cost reductions highly competitive with pure pytket-dqc (26.70\%) while operating approximately $24\times$ faster than DMapS (a nearly two-order-of-magnitude improvement over pytket-dqc's CE and PH variants). Finally, our scalable QIG approach sacrifices explicit multi-gate awareness during partitioning to drastically reduce complexity. Remarkably, QIG + CLA-SABRE ($|E|=20$) still secures a 24.72\% cost reduction while achieving a $140\times$ speed-up against DMapS. This demonstrates that extracting global structural properties via lightweight graph partitioning is an effective, computationally sustainable alternative to heavy hypergraph methods.

Regarding our standalone SABRE heuristics on these \textbf{unstructured} circuits, injecting DQC-specific topological awareness consistently improves cost reductions. Interestingly, indiscriminately expanding the lookahead window, such as to $|E|=100$, degrades performance relative to $|E|=20$ (Table \ref{tab:master_benchmark_results_a}; Fig. \ref{fig:48-qubits}). We deduce that an overly aggressive lookahead penalizes immediate DQC-awareness, as distant gate dependencies overshadow immediately actionable teleportations, a dynamic that is inverted on the 399-qubit topologies, where the expanded lookahead becomes highly advantageous.

Shifting to the \textbf{structured} suite, the necessity of our scalable framework becomes undeniable. On these workloads, standalone SABRE routing struggles significantly against DMapS (Table \ref{tab:master_benchmark_results_a}), underscoring the necessity of an initial partitioning phase. Although their strong transpilation quality on unstructured circuits suggests hypergraph approaches would perform well here, they proved computationally unviable: pure pytket-dqc timed out on 14 of the 20 circuits within the strict 1-hour limit. As a caveat, DMapS required a singular exception to this limit ($\approx2.0$ hours) to complete the Shor's algorithm circuit. This exception was mathematically necessary to establish a valid denominator for our geometric means, but it highlights the scaling limits of current state-of-the-art tools. Conversely, QIG + CLA-SABRE ($|E|=20$) successfully processes the entire suite, securing a 24.01\% average cost reduction against DMapS while operating $253\times$ faster. This critical regime highlights our primary contribution: we effectively address the scaling limits of the state-of-the-art, preserving high transpilation quality while enabling practical runtimes.

\subsubsection{Performance on 399-Qubit Architectures}

% ============================================================
% TABLE III — 399-Qubit, full width (table*, needs the space)
% ============================================================
\begin{table*}[t]
    \centering
    \caption{
        Aggregated Cost Reduction and Runtime Speed-up (Geometric-Mean Ratios): 399-Qubit IBM Flamingo (3\,$\times$\,133), Relative to Def.\ SABRE. Hypergraph methods timed out on the entire suite and are omitted. Bold values are best per column. See footnotes for partial-result superscripts. Method legend matches Table~\ref{tab:master_benchmark_results_a}; (with 3\,h timeout for the 399-qubit architectures).
    }
    \label{tab:master_benchmark_results_b}
    \vspace{-0.15cm}
    \resizebox{\linewidth}{!}{%
        \begin{tabular}{l rr p{0.33em} rr p{1.0em} rr p{0.33em} rr}
            \toprule
            \multirow{2}{*}{\textbf{Algorithm}}
                & \multicolumn{5}{c}{\textbf{All-to-All Topology}}
                &
                & \multicolumn{5}{c}{\textbf{Line Topology}} \\
            \cmidrule(lr){2-6}\cmidrule(lr){8-12}
                & \multicolumn{2}{c}{Unstructured ($7{\times}5{+}1$)}
                &
                & \multicolumn{2}{c}{Structured (16 Circuits)}
                &
                & \multicolumn{2}{c}{Unstructured ($7{\times}5{+}1$)}
                &
                & \multicolumn{2}{c}{Structured (16 Circuits)} \\
            \cmidrule(lr){2-3}\cmidrule(lr){5-6}%
            \cmidrule(lr){8-9}\cmidrule(lr){11-12}
                & \multicolumn{1}{c}{Cost Red.}
                & \multicolumn{1}{c}{Speed-up}
                &
                & \multicolumn{1}{c}{Cost Red.}
                & \multicolumn{1}{c}{Speed-up}
                &
                & \multicolumn{1}{c}{Cost Red.}
                & \multicolumn{1}{c}{Speed-up}
                &
                & \multicolumn{1}{c}{Cost Red.}
                & \multicolumn{1}{c}{Speed-up} \\
            \midrule

            % --- SABRE family ---
            (1,10) SABRE
                & $\mathbf{10.30\%}$ & $0.908\times$
                &
                & $-93.10\%$   & $0.764\times$
                &
                & $5.92\%$     & $0.881\times$
                &
                & $-72.64\%$   & $0.763\times$ \\
            CLA ($|E|{=}20$)
                & $-127.79\%$  & $0.095\times$
                &
                & $-106.95\%$  & $0.350\times$
                &
                & $-261.50\%$  & $0.032\times$
                &
                & $-43.47\%$   & $0.184\times$ \\
            CLA ($|E|{=}100$)
                & $-52.62\%$   & $0.053\times$
                &
                & $-58.43\%$   & $0.162\times$
                &
                & $-143.70\%$  & $0.032\times$
                &
                & $-58.43\%$   & $0.143\times$ \\
            \midrule

            % --- QIG family ---
            QIG+Def.\ SABRE
                & $0.35\%$     & $\mathbf{1.029\times}$
                &
                & $46.99\%$    & $\mathbf{1.051\times}$
                &
                & $1.68\%$     & $\mathbf{1.182\times}$
                &
                & $48.41\%$    & $1.264\times$ \\
            QIG+(1,10) SABRE
                & $10.07\%$    & $0.899\times$
                &
                & $40.33\%$    & $0.977\times$
                &
                & $\mathbf{7.21\%}$  & $1.073\times$
                &
                & $43.35\%$    & $\mathbf{1.269\times}$ \\
            QIG+CLA ($|E|{=}20$)
                & $-126.62\%$  & $0.090\times$
                &
                & $47.64\%$    & $0.734\times$
                &
                & $-243.71\%$  & $0.038\times$
                &
                & $48.80\%$    & $0.858\times$ \\
            QIG+CLA ($|E|{=}100$)
                & $-52.62\%$   & $0.053\times$
                &
                & $\mathbf{51.94\%}$ & $0.331\times$
                &
                & $-134.55\%$  & $0.035\times$
                &
                & $\mathbf{51.96\%}$ & $0.373\times$ \\
            \midrule

            % --- DMapS isolated at bottom ---
            DMapS
                & T.O. & T.O.
                &
                & $44.90\%$\textsuperscript{d} & $0.002\times$\textsuperscript{d}
                &
                & T.O. & T.O.
                &
                & $47.43\%$\textsuperscript{d} & $0.002\times$\textsuperscript{d} \\
            \bottomrule
        \end{tabular}
    }%

    \vspace{0.15cm}
    \raggedright
    \scriptsize
    \textsuperscript{d} Subset (14/16); 3-hour timeout on densest circuits (HHL-385, QAOA-398).
    \vspace{-1.75em}
\end{table*}

% ============================================================
% FIG. 3 — Large Benchmarks (399-Qubit), portrait figure* (full width)
% ============================================================
\begin{figure*}[t]
    \centering
    \includegraphics[width=0.95\textwidth]{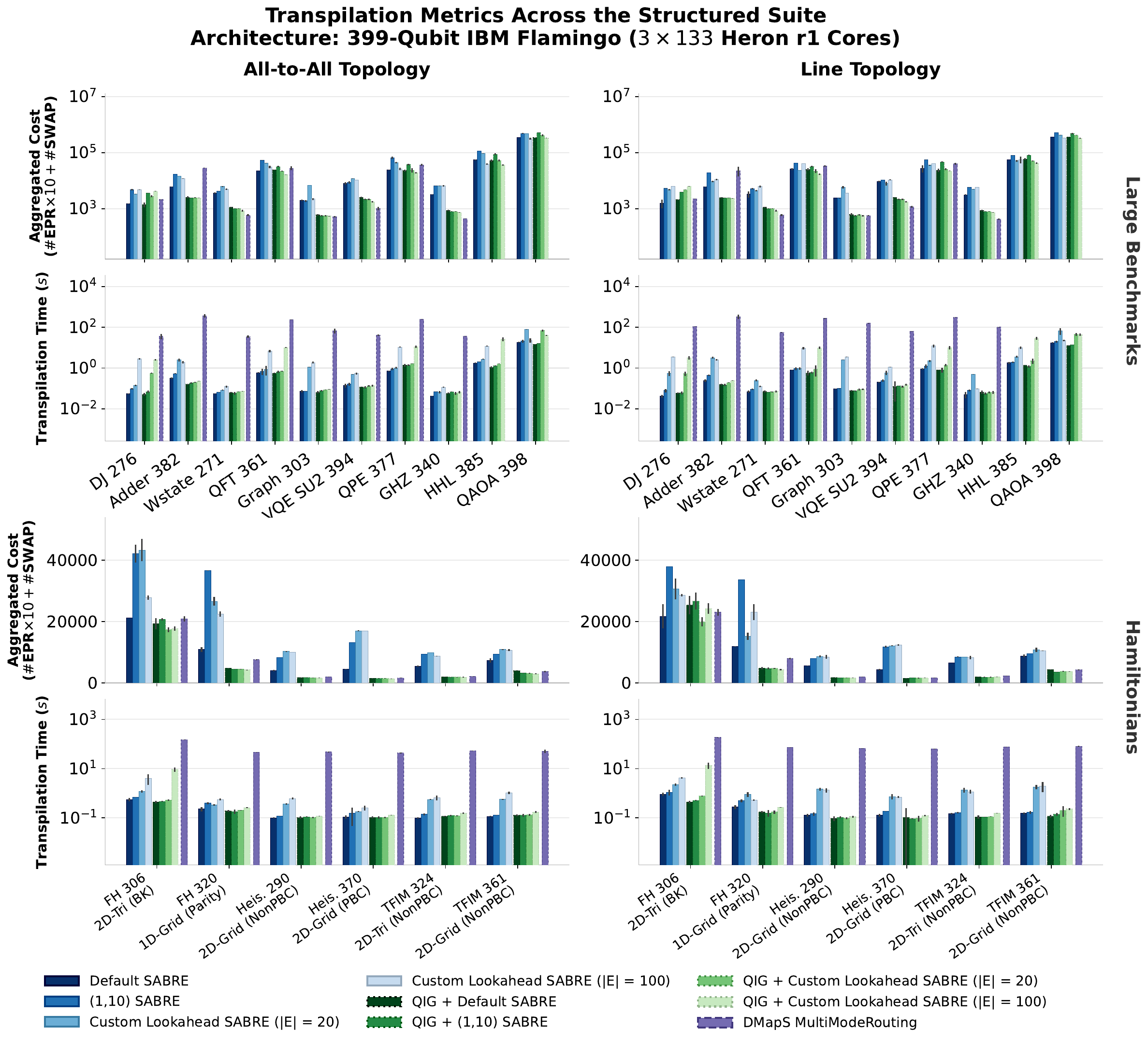}
    \vspace{-0.55em}
    \caption{Results for the structured suite on the 399-Qubit IBM Flamingo architecture (all-to-all \textit{vs.} line topologies). Evaluated metrics: aggregated cost and total runtime (3-hour timeout). Omitted bars reflect timeouts: DMapS exceeded the limit on the densest circuits (HHL-$385$, QAOA-$398$), while hypergraph approaches (red/orange tones) timed out across the entire suite. The color code matches Fig.~\ref{fig:48-qubits}. \textbf{Note:} ``Default SABRE'' is our DQC-adapted SABRE baseline (Sec.~\ref{SecIV:SABRE_Adaptations_for_DQC}).}
    \label{fig:399-qubits}
    \vspace{-1.55em}
\end{figure*}

We now expand our analysis to the IBM Flamingo architectures. Average metrics are detailed in Table \ref{tab:master_benchmark_results_b}, and Fig. \ref{fig:399-qubits} illustrates the circuit-level results for the structured suite across both the all-to-all and line topologies (unstructured circuits are omitted from these plots for visual clarity).

At this scale, the existing state-of-the-art exhibits prohibitive runtime overheads. The hypergraph-based methods exceeded the 3-hour timeout across \textit{all} evaluated circuits (and are consequently omitted from Table \ref{tab:master_benchmark_results_b} and Fig. \ref{fig:399-qubits}). Similarly, DMapS timed out on the entirety of the unstructured suite, as well as on the densest structured benchmarks (HHL, QAOA). Consequently, to establish a viable evaluation denominator, we pivot our reference baseline to Default SABRE (our DQC-adapted variant defined earlier).

For \textbf{structured} workloads on the \textbf{all-to-all} topology, our QIG-based methods bridge the scalability gap by consistently outperforming DMapS on its completed subset with immense computational agility. Specifically, QIG + CLA-SABRE ($|E|=100$) achieves a 51.94\% cost reduction against the Default SABRE baseline, surpassing the 44.90\% reduction achieved by DMapS on its completed subset. Furthermore, it accomplishes this while operating $166\times$ faster than DMapS. Alternatively, QIG + Default SABRE provides a 46.99\% cost reduction while reaching speed-ups exceeding $525\times$ against DMapS. The ability to choose between these distinct routing variants provides valuable flexibility, allowing the prioritization of either superior transpilation quality or rapid compilation times.

For large \textbf{unstructured} circuits on the \textbf{all-to-all} topology, our analysis is restricted to our own contributions due to complete state-of-the-art (DMapS and pytket-dqc) timeouts. Here, adding QIG pre-processing does not significantly alter the results, rendering any QIG + SABRE variant essentially equivalent in performance to its standalone SABRE counterpart (Table \ref{tab:master_benchmark_results_b}). This phenomenon is tied to ``structural homogenization'': large, dense random circuits produce highly uniform interaction graphs with no meaningful clusters to extract. Consequently, QIG partitioning aligns closely with the random initial layouts generated natively by the pure SABRE variants, ultimately yielding similar transpilation quality to them. This contrasts with the 48-qubit scale, where smaller, sparser random circuits still inadvertently contain localized interaction clusters that QIG can exploit. Crucially, this confirms our methodology's viability as a universal compilation tool: it provides substantial routing advantages for structured algorithms, while incurring no performance penalty when applied to highly randomized circuits.

Finally, evaluating the \textbf{restricted line} topology demonstrates our algorithms' generalizability. Transitioning from the all-to-all IBM Flamingo architecture to a line topology results in a moderate degradation of transpilation quality for the unstructured suite (Table \ref{tab:master_benchmark_results_b}), as the reduction in available inter-core links naturally forces higher EPR pair consumption in random circuits. Structured-suite transpilation quality, meanwhile, remains largely consistent with the all-to-all case, indicating our algorithms seamlessly adapt to heterogeneous topologies without requiring manual retuning.

\vspace{-0.15em}
\section{Conclusions and Future Work}
\label{SecVII:Conclusions_and_Future_Work}
\vspace{-0.25em}

This work demonstrates that coupling lightweight quantum interaction graph (QIG) partitioning with DQC-adapted SABRE heuristics provides a highly performant and scalable framework for distributed quantum transpilation. We observe that while computationally intensive hypergraph-based approaches (pytket-dqc \cite{andresmartinez2024pytketdqc}) can yield slightly superior cost reductions on small unstructured circuits, they scale poorly, timing out on the majority ($14/20$) of structured benchmarks even within a modest $48$-qubit ($3~\text{QPUs} \times 16~\text{qubits}$) architecture. Similarly, the current state-of-the-art framework, DMapS \cite{luo2025dmaps}, encounters severe scaling bottlenecks, often timing out on dense workloads targeted for the $399$-qubit ($3 \times 133$) hardware. Our methodology effectively resolves these critical limitations: it remains highly competitive on smaller architectures, achieving geometric mean aggregated cost reductions of up to $24.72\%$ and $24.01\%$ on $48$-qubit unstructured and structured benchmark suites, respectively, relative to DMapS. Furthermore, it delivers these results alongside runtime speed-ups of approximately $140\times$ for unstructured and $>250\times$ for structured workloads compared to DMapS. Crucially, our approach sustains this routing quality at the utility scale. On the $399$-qubit IBM Flamingo architectures, where current state-of-the-art methods (DMapS and pytket-dqc) struggle substantially or time out entirely, it delivers nearly $52\%$ cost reductions against a DQC-adapted SABRE baseline, establishing a computationally sustainable path for transpiling complex workloads onto emerging distributed hardware.

Looking ahead, we identify several key directions for future work. First, we intend to expand our evaluation to more complex, heterogeneous topologies, including ring, star, hypercube, and 2D-torus networks. Such distributed architectures become increasingly relevant as we move toward the fault-tolerant era \cite{swaroop2025universaladaptersquantumldpc, Bravyi2024}. Understanding how our QIG-based partitioning must evolve to accommodate the unique constraints of error-corrected QPUs, where the cost of logical-level routing is not trivially defined, will be critical. Second, we seek to bridge the gap between simple graph partitioning and more informative hypergraph models. Future work should investigate scalable hypergraph partitioning schemes capable of exploiting EJPP protocols in a more computationally efficient manner. Complementary to this, we aim to explore further SABRE modifications, incorporating dynamic ``attraction'' metrics \cite{escofet2023hungarian, baker2020fgproee} to better exploit high-density interaction clusters.

Finally, a promising direction is adapting our framework to dynamically select between quantum gate and state teleportation primitives. Currently, our QIG-based methods manage remote operations through heuristic inter-core SWAP insertions (implemented via sequential quantum gate teleportation), without explicitly scheduling EJPP processes or considering state teleportation. Future work will extend our approach to explicitly utilize multi-gate teleportation (via EJPP) when possible, and state teleportation when more favorable. An ideal extension would intelligently coordinate these primitives based on link usage and available capacity, providing a truly end-to-end solution for the next generation of distributed quantum systems.

\bibliographystyle{IEEEtran}
\bibliography{IEEEabrv,paper}

\end{document}